\documentclass[12pt]{article}
\usepackage{amsmath, amssymb}
\usepackage{graphics,graphicx}
\usepackage{epstopdf}
\usepackage{epsfig,comment}

\newcommand{\lsim}{\,\raise.3ex\hbox{$<$\kern-.75em\lower1ex\hbox{$\sim$}}\,}
\newcommand{\gsim}{\,\raise.3ex\hbox{$>$\kern-.75em\lower1ex\hbox{$\sim$}}\,}

\newcommand{\EE}{\mathcal{E}}

\newcommand{\OO}{\mathcal{O}}

\newcommand{\GeV}{\text{GeV}}

\newcommand{\eff}{{\text{eff}}}

\begin{document}

\begin{titlepage}
\renewcommand{\thefootnote}{\fnsymbol{footnote}}
\setcounter{footnote}{0}
\begin{flushright}
SLAC-PUB-11140\\
hep-ph/0504210\\
\end{flushright}
\vskip 2cm
\begin{center}
{\large\bf Limits on Split Supersymmetry from Gluino Cosmology}
\vskip 1cm {\normalsize Asimina Arvanitaki$^{1}$, Chad Davis$^{1}$,
Peter W. Graham$^{1}$, Aaron Pierce$^{2,1}$\footnote{The work of AP
is supported by the U.S. Department of Energy under contract number
DE-AC02-76SF00515.}, Jay G. Wacker$^{1}$\footnote{JGW is supported
by National Science
Foundation Grant PHY-9870115 and by the Stanford Institute for Theoretical Physics.}\\
\vskip 0.5cm
1. Institute for Theoretical Physics\\
   Stanford University\\
   Stanford, CA 94305\\
\vskip .1in
2. Theory Group \\
   Stanford Linear Accelerator Center\\
   Menlo Park, CA 94025\\
\vskip .1in
}
\end{center}

\vskip .5cm

\begin{abstract}
An upper limit on the masses of scalar superpartners in split
supersymmetry is found by considering cosmological constraints on
long-lived gluinos. Over most of parameter space, the most stringent
constraint comes from big bang nucleosynthesis.  A TeV mass gluino
must have a lifetime of less than 100 seconds to avoid altering the
abundances of D and ${}^6 \text{Li}$. This sets an upper limit on
the supersymmetry breaking scale $m_{S}$ of $10^9$ GeV.
\end{abstract}

\end{titlepage}

\section{Introduction}
\renewcommand{\thefootnote}{\arabic{footnote}}
\setcounter{footnote}{0} In split supersymmetry
\cite{Arkani-Hamed:2004fb, Giudice:2004tc} (see also
\cite{Wells:2003tf}), the superpartner spectrum differs drastically
from that of traditional weak scale supersymmetry.  The fermionic
superpartners are present at the weak scale, while the scalar
superpartners have much larger masses, at a scale $m_{S}$. This
represents a new scale, which is a priori undetermined. Verification
of split supersymmetry will require not only the detection of the
new TeV-mass states at colliders, but also the observation of
indirect effects of the full supersymmetry present above $m_{S}$.

Most probes of $m_{S}$ are only logarithmically sensitive to this
scale \cite{Cheng:1997sq}.  For example, couplings whose values are
determined by supersymmetry deviate via renormalization group flow once
supersymmetry is broken.  The possibility of using such effects to
determine $m_{S}$ in split supersymmetry was explored in
\cite{Arkani-Hamed:2004fb, Arvanitaki:2004eu}. There is one
observable that has power-law sensitivity to the supersymmetry
breaking scale--the gluino lifetime.  Thus, measurement of the
gluino lifetime would allow a precise determination of $m_{S}$. This
motivates us to look at cosmological constraints on the gluino lifetime.
For large scalar masses, the gluino can become long-lived.
The possibility that the successful predictions of
standard big bang cosmology might be altered by a long-lived gluino
presents an opportunity to constrain $m_{S}$.

The effects of the late decaying gluinos depend on their
annihilation cross section, which determines their relic abundance.
The calculation of this cross section is complicated by the strong
interactions of the gluino\cite{Baer:1998pg,PastRelic}. The relic
abundance calculation requires an estimate of non-perturbative
annihilation processes after the QCD phase transition.  We discuss
the computation of the relic abundance of the gluinos in Section
\ref{Sec: Relic}.

In Section \ref{Sec: Limits}, we use the results of this relic abundance
calculation to place limits on the gluino lifetime.  A particularly
strong constraint comes from Big Bang Nucleosynthesis (BBN).  We
find that for TeV mass gluinos, the lifetime is limited to be less
than 100 seconds.  Finally, we relate these lifetime bounds to
bounds on the scale $m_{S}$.

\section{Annihilation of Gluinos}
\label{Sec: Relic}

The relic density of gluinos prior to their decay is determined by
evolving the Boltzmann equation.  This determines the freeze-out temperature
where  the expansion rate of the universe is balanced against the
annihilation rate of the gluinos. There are two
regimes of annihilation that we will consider separately. The first
era is before the QCD phase transition when there are free gluinos
in a QCD plasma. The second era is after the QCD phase transition
when the gluinos have become confined in color neutral $R$-hadrons.  The
annihilation cross section in this second period can, in principle,
be much higher than in the first, thus leading to a second period
of annihilation.

\subsection{Perturbative}

The perturbative annihilation of gluinos is well known
\cite{Baer:1998pg} and at low velocities the cross section is given by
\begin{eqnarray}
\sigma(\tilde{g}\tilde{g} \rightarrow g g) = \frac{27 \pi \alpha_s^2}{64 m_{\tilde{g}}^2 v} + \OO(v^0)\hspace{0.5in}
\sigma(\tilde{g}\tilde{g} \rightarrow q \bar{q  }) = \frac{9 N_f^{\eff}\pi \alpha_s^2}{64 m_{\tilde{g}}^2 v} + \OO(v^0),
\end{eqnarray}
where  $N_f^{\eff}$ is the effective number of flavors of quarks and
includes a phase space suppression for the top quark.  We use the
full expression for the annihilation cross section  \cite{Baer:1998pg}
in our numerical calculations.

Because the velocities of the gluinos are small at freeze-out, there
is a Sommerfeld enhancement due to $t$-channel exchange of multiple
gluons \cite{sommerfeld}.    Each loop of gluons gives a $\pi
\alpha_s/v$ enhancement but can be resummed into an overall
enhancement 
\begin{eqnarray}
\EE_{gg} = \frac{ \frac{1}{2} \pi \alpha_s/v}{1 - \exp(\frac{1}{2} \pi \alpha_s/v)}
\hspace{0.5in}
\EE_{q\bar{q}} = \frac{ \frac{3}{2} \pi \alpha_s/v}{1 - \exp(\frac{3}{2} \pi \alpha_s/v)}.
\end{eqnarray}
Taking the enhancement into account, the cross section becomes 
\begin{eqnarray}
\sigma^{\text{tot}}(\tilde{g}\tilde{g} \rightarrow g g) = \frac{27 \pi^2 \alpha_s^3}{128 m_{\tilde{g}}^2 v^2} +\OO(v^{-1})
\hspace{0.5in}
\sigma^{\text{tot}}(\tilde{g}\tilde{g} \rightarrow q \bar{q}) = \frac{27 N_f^{\eff}\pi^2 \alpha_s^3}{128 m_{\tilde{g}}^2 v^2} + \OO(v^{-1}).
\label{Eqn: Enhanced}
\end{eqnarray}
Because the Sommerfeld enhancement is a long-distance effect, one
might worry that this effect is suppressed due to thermal effects
from the QCD plasma.    However, the effective mass of a gluon in a
plasma is $\alpha_s T$, while the typical momentum associated with
the Sommerfeld enhancement loops is $\alpha_s m_{\tilde{g}} v\sim
\alpha_s \sqrt{m_{\tilde{g}} T}\gg \alpha_s T$.  Thus, the
Sommerfeld enhancement remains effective.

Sommerfeld enhancement can be interpreted as annihilation through
off-shell bound states of gluinos.  One could also ask whether bound
states are formed on-shell.  This requires radiating away energy.
This will suppress the formation of such states by perturbative
powers of $\alpha$.  We will revisit bound state formation in a
non-perturbative regime in the next section.

\subsection{Non-Perturbative}

After the QCD phase transition, the gluinos hadronize and form
R-hadrons.  Very little is certain about the spectroscopy, quantum
numbers, or couplings of these strongly interacting particles.
In order to have a reliable calculation of the relic density of
gluinos, we must nevertheless estimate the annihilation cross section of
R-hadrons.   Estimates of this cross section have varied widely, due
to different assumptions about the relevant hadronic physics. A
commonly made assumption is that the annihilation cross section goes
as  $\pi R^2 \simeq 1/m_{\pi}^2$, where $R$ is the geometric size of
the R-hadron.  This cross section would be roughly dozens of
millibarns, much larger than the perturbative
annihilation cross section. In this section we argue that this is
not the case.  There is a huge separation of scales
between the gluino mass and the hadronization scale. This makes it
hard to exchange momentum efficiently between the QCD cloud and the
gluino core.  Thus, we expect the light QCD degrees of freedom to decouple
from the annihilation process.

Geometric ($\sigma \sim R^2$) cross sections for $R$-hadron
scattering are applicable in the case of  $q^{2} <
\Lambda_{QCD}^{2}$.  Such processes do not probe the interior
structure of the hadron because they transfer low momentum, and the
hadron appears like a solid object.  On the other hand, gluino
annihilation is a large momentum transfer process and probes the
interior structure of the hadron.  The partonic picture
of the hadron becomes relevant, and we
expect the annihilation cross section is set by the size of the
parton, $\sigma_{ann}  \sim m_{\tilde{g}}^{-2}$.

To explore this further, we can rely on simple quantum mechanical
scattering arguments. Despite the strong dynamics, the inelastic
$R$-hadron cross section must still satisfy partial wave unitarity
\cite{Griest:1989wd} 
\begin{eqnarray}
\sigma_j^{inel} \le \frac{ \pi}{m_{\tilde{g}}^2 v^2} (2j +1).
\label{Eqn: PW}
\end{eqnarray}
Notice that the cross section scales as the de Broglie wavelength of
the gluino rather than a geometric quantity.  However, this does
not prove that the annihilation cross section scales as
$(m_{\tilde{g}} v)^{-2}$.  If many partial wave cross sections
contribute, they can add up to a geometric cross section.
For example, if all partial waves up to  $j_{\text{max}} =
m_{\tilde{g}} v R$ contribute, where $R$ is the size of the
R-hadron, then Eqn.~\ref{Eqn: PW} sums to give a geometric cross
section, ~$R^{2}$. For this to occur, high angular momentum partial waves must
contribute (i.e. $j\simeq 30$ for a TeV scale gluino at temperatures
around $\Lambda_{\text{QCD}}$). However, for these high angular
momentum modes to lead to annihilation, the gluinos must effectively
tunnel through a large barrier.  This means that direct annihilation
is exponentially suppressed for high $j$. A significant annihilation cross
section through larger angular momentum partial waves typically
requires either that the object be of uniform density (i.e. like a
macroscopic composite system) or that there are high angular
momentum bound states that contribute to the cross section.

One potential way to avoid the tunneling suppression at large
angular momenta would be via the exchange of a high $j$ QCD
resonance. In this case, the resonance itself would carry the
angular momentum, and could lead to immediate annihilation.  While
this might be plausible if the gluinos had a mass in the GeV range,
it does not seem plausible for TeV mass gluinos-- there are no
sharp QCD resonances well above the GeV scale.  If such a resonance were
to exist, it would be extraordinarily broad, and would therefore not
lead to rapid resonant annihilation.

A second way to bleed off angular momentum would be by radiating.
For the relevant large angular momentum states, we can use a
semi-classical treatment where radiation is caused by accelerating a
particle.  This acceleration could be caused if a QCD string formed
between the two gluinos.     In this case, the radiation may be
described via the Larmor formula where the power radiated is 
\begin{equation}
P \sim a^{2} \sim \frac{1}{(\alpha^{\prime} m_{\tilde{g}})^{2}},
\end{equation}
where $\alpha^{\prime}$ is the tension of the exchanged QCD string.
The total energy radiated can be estimated by multiplying this power
by the crossing time $t_{cross} \sim R/v$, where $v$ represents the
relative velocity of the two hadrons.  For $T \sim \Lambda_{QCD}$,
we find
\begin{equation}
E_{rad} \sim \left(\frac{\Lambda_{QCD}}{m_{\tilde{g}}}\right)^{3/2} \Lambda_{QCD}.
\end{equation}
Thus, radiation from the gluinos is small, in fact, much less than
the mass gap to the lightest possible state that could be radiated
(the pion).  This agrees with the intuition that heavy objects do
not radiate.  We should note that we could have applied a similar
Larmor argument prior to the QCD phase transition.  In this case,
the relevant force is not due to a QCD string, but rather to a QCD
Coulomb potential.  In this case, the radiation will be further
suppressed by perturbative powers of $\alpha_{s}$, again arguing
against a large rate for the formation of bound states.  Light QCD
degrees of freedom do not carry the momentum or angular momentum of
the system.  Radiation from the cloud therefore is not able to
reduce the relative angular momentum of the heavy gluinos, so they
remain incapable of direct annihilation.

\subsection{Relic Abundance Summary}

\begin{figure}[t]
\label{Fig: Relic}
\begin{center}
\epsfig{file=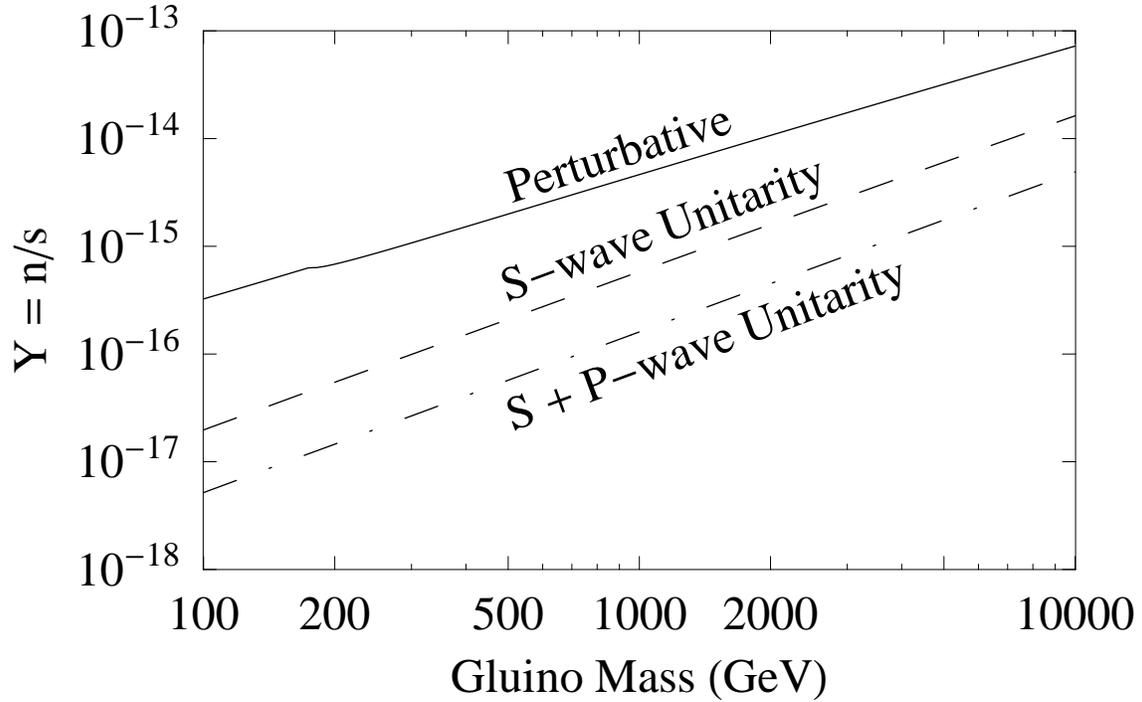, width=6.0in} \caption{Gluino
abundance per co-moving volume as a function of mass.  Three curves
are shown.  In the first (solid), the annihilation cross section is
assumed to be simply given by the perturbative cross section of
Eqn.~\ref{Eqn: Enhanced}.  The other curves correspond to a cross
section that saturates $s$-wave (dashed) and $s$-wave plus $p$-wave
unitarity (dot-dashed).}
\end{center}
\end{figure}

In summary, we estimate
the total annihilation cross section as follows: for $T >
\Lambda_{QCD}$, we simply utilize the perturbative cross section for
gluinos \cite{Baer:1998pg}; for $T < \Lambda_{QCD}$, we allow for the
possibility of an increased cross section.  However, because we have argued that high angular momentum states do not
significantly contribute to the annihilation of gluinos, we do not allow for an arbitrarily large cross section at this stage.  We expect that the
annihilation is conservatively given by a cross section that
saturates s-wave unitarity.  We take $\Lambda_{QCD} =200$ MeV as the point where the cross section changes.

This cross section is thermally
averaged \cite{Gondolo:1990dk} and used in the numerical integration
of the Boltzmann equation.  We treat the QCD phase transition as
second order although this is not expected to significantly affect
the results \cite{Srednicki:1988ce}.

For completeness, in Fig.~2, we show the relic abundance for three
cases. First, a solid curve that shows the relic abundance, assuming
all annihilation is perturbative.  Second, a dashed curve that
incorporates the turn on of a cross section that saturates  $s$-wave
unitarity.  Finally, a dot-dashed curve that has a cross section
that saturates $s$-wave plus $p$-wave unitarity, which we view as an
even more conservative assumption.

\section{Limits}
\label{Sec: Limits}

In this section we discuss the effects of the decays of the relic
gluinos.  Depending upon the lifetime of the gluino, these decays
can disturb the predictions of BBN, distort the Cosmic Microwave
Background Radiation (CMBR), or show up in the diffuse gamma ray
background. If the gluinos have a lifetime of order the age of the
universe, they can potentially be observed in experiments searching for
heavy exotic nuclei.  We now discuss each of these constraints.

First, we consider the era of BBN.  These constraints are a function
of the gluino destructive power, $\xi \equiv E_{vis} Y$. Here
$E_{vis}$ is the energy of the gluino decay products which are
deposited in the thermal bath, and $Y \equiv n/s$ represents the
number density of gluinos per co-moving volume. To determine
$E_{vis}$, we assume that half of the energy of a decaying gluino
goes into the neutralino, while the other half is distributed
between hadrons.  At the time of the gluino decay, the neutralino
mean free path is large, and thus the energy in the neutralino is
simply carried away; it is not ``visible''. That is to say, we are
taking the visible energy $E_{vis} = m_{\tilde{g}}/2$.  If the lightest
supersymmetric particle has a mass close to that of the gluino, the limits
are weakened for a given abundance.

Constraints arising from BBN on the lifetime of a hadronically
decaying particle as a function of its destructive power were
described in \cite{Kawasaki:2004qu}.  We take the most conservative
D/H and ${}^4 \text{He}$ limits.  We are interested in the limits as
a function of the gluino mass, but they only present results for
three masses (100 GeV, 1 TeV, and 10 TeV). For a fixed lifetime, the
dependence of $\xi$ on $m_{\tilde{g}}$ is roughly power-law, and so
we can use these three points to interpolate to different values of
the gluino mass. This procedure gives bounds on the gluino lifetime
as a function of mass. We then convert lifetime bounds to bounds on
the supersymmetry breaking scale \cite{Toharia:2005gm, Gambino:2005eh}, augmenting the decay rate by the two-body decay rate
when significant. The result is shown in Fig.~2.

We now discuss in more detail the origin of the BBN bounds. Three
BBN bounds are particularly restrictive: D/H, ${}^6 \text{Li/H}$,
and ${}^3 \text{He/D}$.  After BBN has finished, the universe is
composed primarily of hydrogen and helium, with trace amounts of
other elements. A gluino decay releases particles at energies much
higher than the typical kinetic energies of nuclei at that time.  At
times later than about 100 seconds, mesons decay before they
scatter, so the most destructive hadronic decay products are
baryons. Protons produced by gluino decays scatter off of the
background protons and ${}^4 \text{He}$ nuclei. The elastic
scattering increases the energy of background protons, making them
more likely to react with other elements.  The inelastic $\text{p} +
{}^4 \text{He} \rightarrow {}^3 \text{He} + \text{D}$ reaction
increases D/H. If a D scatters off of a background ${}^4 \text{He}$
nucleus, ${}^6 \text{Li}$ can form.  The epoch important for D
formation is around 100 seconds -- this sets the lower edge of the
BBN curve in Fig.~2.

A second process important in BBN is photo-dissociation\cite{Ellis}.  Photons with energies
above 20 MeV can break ${}^4\text{He}$ nuclei into
${}^3\text{He}$ or Tritium (which later decays weakly to
${}^3\text{He}$). This can cause the ${}^3 \text{He/D}$ ratio to
become too large.  However, this process does not become important until
the temperature drops to $\sim1$ keV at $\sim 10^{6}$ seconds
(see, e.g. \cite{Kawasaki:1994sc}).  At
temperatures before this, the 20 MeV photons lose energy efficiently by
scattering off the background photons before they can break apart a nucleus.

\begin{figure}[t]
\label{Fig: Bounds}
\begin{center}
\epsfig{file=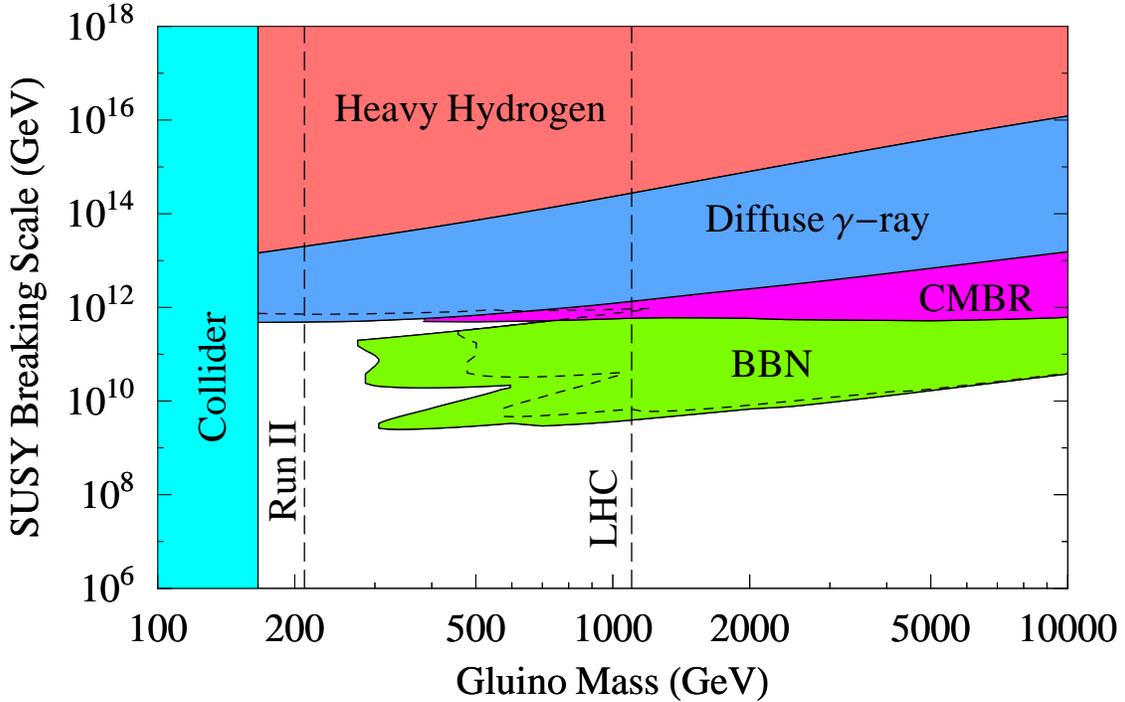, width=6.0in} \caption{Limits on the
supersymmetry breaking scale, $m_{S}$, as a function of the gluino
mass, $m_{\tilde{g}}$.  The bounds are derived assuming a
perturbative cross section in Eqn. ~\ref{Eqn: Enhanced} for
temperatures greater than the QCD phase transition, $T > 200$ MeV.
For $T < 200$ MeV, we assume that the annihilation cross section
saturates $s$-wave unitarity.  Also shown (dashed) are the limits in
the case where the annihilation cross section saturated $s$-wave
plus $p$-wave unitarity.  The shaded regions are excluded.  The
lower edge of the BBN curve arises from the requirement that the
$D/H$ ratio remained undisturbed, and corresponds to a lifetime of
approximately 100 seconds.}
\end{center}
\end{figure}

Gluinos that decay after recombination can give rise to photons that
free-stream to us, and are visible in the diffuse gamma ray
background \cite{Kribs:1996ac}. Photons are produced when gluinos
decay to pions which subsequently decay to gamma rays. Observations
by EGRET\cite{EGRET} and COMPTEL limit the flux of these gamma rays
and thus the relic abundance of gluinos.  A 3-body decay including an
invisible product, in our case the neutralino, was assumed in
\cite{Kribs:1996ac}. This is also the case at hand.  The region excluded
is shown in Fig.~2.  These observations of the diffuse gamma ray
background rule out a gluino with lifetime around the present age of
the universe.

The CMBR can be used to put further limits on the relic abundance of
gluinos with lifetimes up to $\sim 10^{13}$ s \cite{Hu:1993gc,
Feng}. Gluinos that decay during or after the epoch of
thermalization of the CMBR can distort its spectrum and are thus
limited by observations from COBE\cite{COBE}.  To derive these
limits we make the conservative assumption that roughly $10 \%$ of
the energy of the gluino gets transmitted to the CMBR when it
decays, with the remainder carried off by neutrinos or the lightest
supersymmetric particle. The limits derived from these constraints
are shown in Fig.~2.

If the gluino lives to the present day we would expect it to bind
into nuclei, producing anomalously heavy isotopes.  A combination of
time of flight, mass spectrometry, and direct density measurements
places severe limits on the abundance of terrestrial heavy elements
today \cite{Smith:1982qu, Hemmick:1989ns}.  This rules out a gluino
with lifetime greater than the present age of the universe for the
entire gluino mass range considered.  These bounds are weakened if
it is assumed that the heavy elements sink towards the center of the
earth. However, for heavy hydrogen, the equilibrium time constant is
$\gtrsim 10^8$ years and so if the oceans undergo mixing the heavy
hydrogen would be roughly uniformly distributed \cite{Smith:1982qu}.
The bounds derived from these searches are shown in Fig.~2, labeled
``Heavy Hydrogen.''

Gluino properties are already restricted via direct
searches at colliders\cite{Kilian,Hewett:2004nw}. The weakest limits
come in the case where the produced $R$-hadrons are neutral. Then,
they escape the detector, carrying away a substantial fraction of
the event energy.  The gluinos are then observed in mono-jet events,
triggered by the presence of an additional radiated jet. Current
limits on the gluino from Tevatron Run I are $m_{\tilde{g}}
> 170 ~\GeV$ \cite{Hewett:2004nw}.  This bound should increase to
$210 ~\GeV$ if Run II sees nothing, and to $1100 ~\GeV$ at the Large
Hadron Collider (LHC). These bounds are independent of the
supersymmetry breaking scale.  If charged $R$-hadrons are produced
that do not immediately decay, this bound improves.
Searching for gluinos through anomalously slow tracks in the tracking
chambers of both ATLAS and Tevatron were studied in \cite{Kraan,RunII}.
These provide search reaches comparable to the monojet signal and will
be useful as an additional discovery channel.

Finally there is the possibility of seeing gluinos in cosmic rays and
was studied by \cite{Anchordoqui:2004bd}.  If a gluino were seen,
then this would set and lower limit on the susy breaking scale.
While most of the available lifetime and mass ranges are ruled
out by the considerations in this paper, there is a window available at 
low gluino masses for gluinos to be seen in cosmic rays.

\section{Conclusion}

In this note we have explored the limits on the split supersymmetry
parameter space coming from cosmological constraints on a long lived
gluino.   If the gluino were to annihilate with a geometric cross
section after the QCD phase transition, then it would evade most
cosmological constraints.   However, we have argued that this is not
the case, the cross section is more likely set by the de Broglie
wavelength of the gluino.  Then, the cross section is likely bounded
by saturating s-wave unitarity.   Using this cross section, it is
possible to extract useful limits from the CMBR, BBN, and the
diffuse gamma ray background. For gluinos heavier than 300 GeV the
earliest cosmological constraints are from BBN, specifically the D/H
ratio.    This sets an upper bound to the lifetime of roughly 100
seconds and an upper bound on $m_S$ of $10^9$ GeV. For gluinos
lighter than 300 GeV, the earliest constraints arise from
non-observation of diffuse gamma rays and sets an upper limit to the
lifetime of $10^6$ years and corresponds to $m_S < 10^{11}$ GeV. If
collider searches at the LHC find a gluino heavier than 600 GeV,
then the lifetime will be less than 100 seconds. This has
implications for experiments that hope to trap the gluino and
directly measure its lifetime.

\section*{Acknowledgments}
We would like to thank N. Arkani-Hamed, S. Dimopoulos, E. Katz, M.
Peskin, and J. Wells for useful discussions. A previous version of the paper included a calculation of renormalization of the four Fermi operators associated with the three-body decays of the gluino due to QCD effects. We thank P. Gambino, G. F. Giudice and P. Slavich 
for pointing out an error in our calculation of the gluino lifetime that appeared in the original
version of this paper \cite{Gambino:2005eh}. 
P.W.G. is supported by the National Defense Science and Engineering Graduate Fellowship.
C.D. is the Mellam Family Graduate Fellow.


\providecommand{\href}[2]{#2}\begingroup\raggedright

\endgroup

\end{document}